\documentclass[final,5p,times,twocolumn]{elsarticle}

\usepackage{graphicx}
\usepackage{dcolumn}
\usepackage{bm}
\usepackage{color}
\usepackage{ifthen}
\usepackage[colorlinks=true, linktocpage=true, allcolors=blue]{hyperref}
\usepackage{amsmath}

\def\urlprefix{}

\biboptions{square,numbers,sort&compress}
\journal{Journal of Physics: Conference Series (SCES2016)}

\begin{document}
\begin{frontmatter}
\title{Ce\,3$p$ hard x-ray photoelectron spectroscopy study of\\ the topological Kondo insulator CeRu$_4$Sn$_6$}

\address[Cologne]{Institute of Physics II, University of Cologne, Z{\"u}lpicher Stra{\ss}e 77, 50937, K\"{o}ln, Germany}
\address[Dresden]{Max Planck Institute for Chemical Physics of Solids, N{\"o}thnitzerstra{\ss}e 40, 01187 Dresden, Germany}
\address[SOLEIL]{Synchrotron SOLEIL, L'Orme des Merisiers, Saint-Aubin PB48, 91192 Gif-sur-Yvette, France}
\address[Hsinchu]{National Synchrotron Radiation Reasearch Center, Hsinchu, Taiwan}
\address[Vienna]{Institute of Solid State Physics, Vienna University of Technology, Austria}
\address[Hiroshima]{Department of Quantum Matter, ADSM Hiroshima University, Japan}

\author[Cologne]{M.~Sundermann}
\author[Cologne]{K.~Chen}
\author[Dresden,SOLEIL]{Y.~Utsumi}
\author[Hsinchu]{Y.-H.~Wu}
\author[Hsinchu]{K.-D.~Tsuei}
\author[Vienna]{J.~Haenel}
\author[Vienna]{A.~Prokofiev}
\author[Vienna]{S.~Paschen}
\author[Hiroshima]{A.~Tanaka}
\author[Dresden]{L.\,H.~Tjeng}
\author[Cologne]{A.~Severing}

\begin{abstract}
Bulk sensitive hard x-ray photoelectron spectroscopy data of the Ce\,3$p$ core level of CeRu$_4$Sn$_6$ are presented.
Using a combination of full multiplet and configuration iteration model we were able to obtain an accurate lineshape analysis of the data, thereby taking into account correlations for the strong plasmon intensities.
We conclude that CeRu$_4$Sn$_6$ is a moderately mixed valence compound with a weight of 8\,\% for the Ce\,$f^0$ configuration in the ground state.
\end{abstract}

\begin{keyword}
topological Kondo insulator \sep valence \sep hard x-ray photoelectron spectroscopy \sep full multiplet \sep single impurity Anderson model 
\end{keyword}

\end{frontmatter}

\section{Introduction}
Kondo insulators are presently experiencing a renaissance since they are promising candidates for correlated materials with nontrivial topology\,\cite{Dzero2010,Takimoto2011,Dzero2012,Dzero2013,Alexandrov2013,Lu2013,Dzero2016}.
Such predictions caused a flurry of experimental investigations on SmB$_6$ (see Ref.\,\cite{Dzero2016} and references therein) and more recently we suggested that also CeRu$_4$Sn$_6$ might be a good candidate for a strongly correlated material with topologically protected surface states\,\cite{Sundermann2015}.

CeRu$_4$Sn$_6$ is a tetragonal Kondo insulator where the hybridization of $f$ and conduction electrons leads to the opening of a gap\,\cite{Das1992,Strydom2005,Paschen2010,Bruning2010,Wissgott2016} at low temperature.
Accordingly, the resistivity rises as temperature decreases, but saturates below $T$$\approx$10\,K.
Our recent spectroscopic investigation of CeRu$_4$Sn$_6$ showed that it has a large Kondo temperature of about 170\,K and a $\Gamma_6$ ground state crystal-field symmetry ($J_z$\,=\,$\pm$1/2)\,\cite{Sundermann2015} which is the favorable symmetry for strong hybridization and opening of a gap according to Dzero \textsl{et al.}\,\cite{Dzero2012,Dzero2013}.
The opening of a gap in the presence of the $J_z$\,=\,1/2 symmetry is supported by our band structure calculations and we inferred that CeRu$_4$Sn$_6$ must be topologically non trivial. 

In that work\,\cite{Sundermann2015} we had deduced the Kondo temperature from the temperature evolution of the $4f^0$ spectral weight in the $L_3$-edge absorption data where we had detected the L$_{\alpha1}$ emission line.
A difficulty of the analysis of the $L$-edge data in general is the assignment of the spectral weights, even when measured in the life time reduced partial fluorescence yield mode, since the line shape reflects the empty 5$d$ density of states in the presence of the core hole.
This uncertainty does not affect the temperature dependence but the absolute numbers for the spectral weights. In Ref.\,\cite{Sundermann2015} we have been fitting empirical, identical line shapes to each $f^n$ configuration and found an absolute $f^0$ occupation of $\approx$\,6\%.
Hard x-ray photoelectron spectroscopy (HAXPES) is a bulk sensitive technique\,\cite{Tanuma2011,Braicovich1997,Laubschat1990,Sekiyama2000} and its data can be analyzed more quantitatively\,\cite{Allen1986,Hufner1992} than that of x-ray absorption.

In cerium based Kondo systems the electron in the $f$ shell hybridizes with the conduction electrons so that the resulting hybridized ground state is a mixed state that can be written as $\Psi_\text{GS}$\,=\,$c_0$ $|\,f^0 \rangle$\,+\,$c_1$\,$|\,f^1\,\underline{v}\,\rangle$\,+\,$c_2$\,$|\,f^2 \underline{\underline{v}}\,\rangle\,$ where the main component is given by the trivalent $f^1$ configuration with additional contributions of the divalent and tetravalent states ($f^2$ and $f^0$) with weights $w_n$\,$\equiv$\,$|c_n|^2$ ($n$\,=\,0,1,2).
Here $\underline{v}$ $\left(\underline{\underline{v}}\right)$ denotes one (two) hole in the valence states.
A core hole acts differently on the respective $f^n$ configurations so that they are split in a PES and XAS experiment and the spectra exhibit three spectral weights that are related to the initial state $f^n$ admixtures. However, due to strong final state effects \textsl{spectral} weights and $f^n$ weights $w_n$ in the ground state are not related 1:1. Here a configuration interaction model is required for converting between the two.
Plasmons may complicate the situation when they cover up or distort the $f^n$ spectral weights.
In such cases the spectra have to be corrected for plasmons by performing a full multiplet calculation that contains the plasmons as part of the lineshape (see Analysis).
Below we present HAXPES data of CeRu$_4$Sn$_6$ at T\,$\le$ 90\,K that have been analyzed quantitatively with the combination of a full multiplet and configuration interaction calculation. 

\section{Experiment \& Analysis}

\begin{figure}[h]
\centering
\includegraphics[height=26pc]{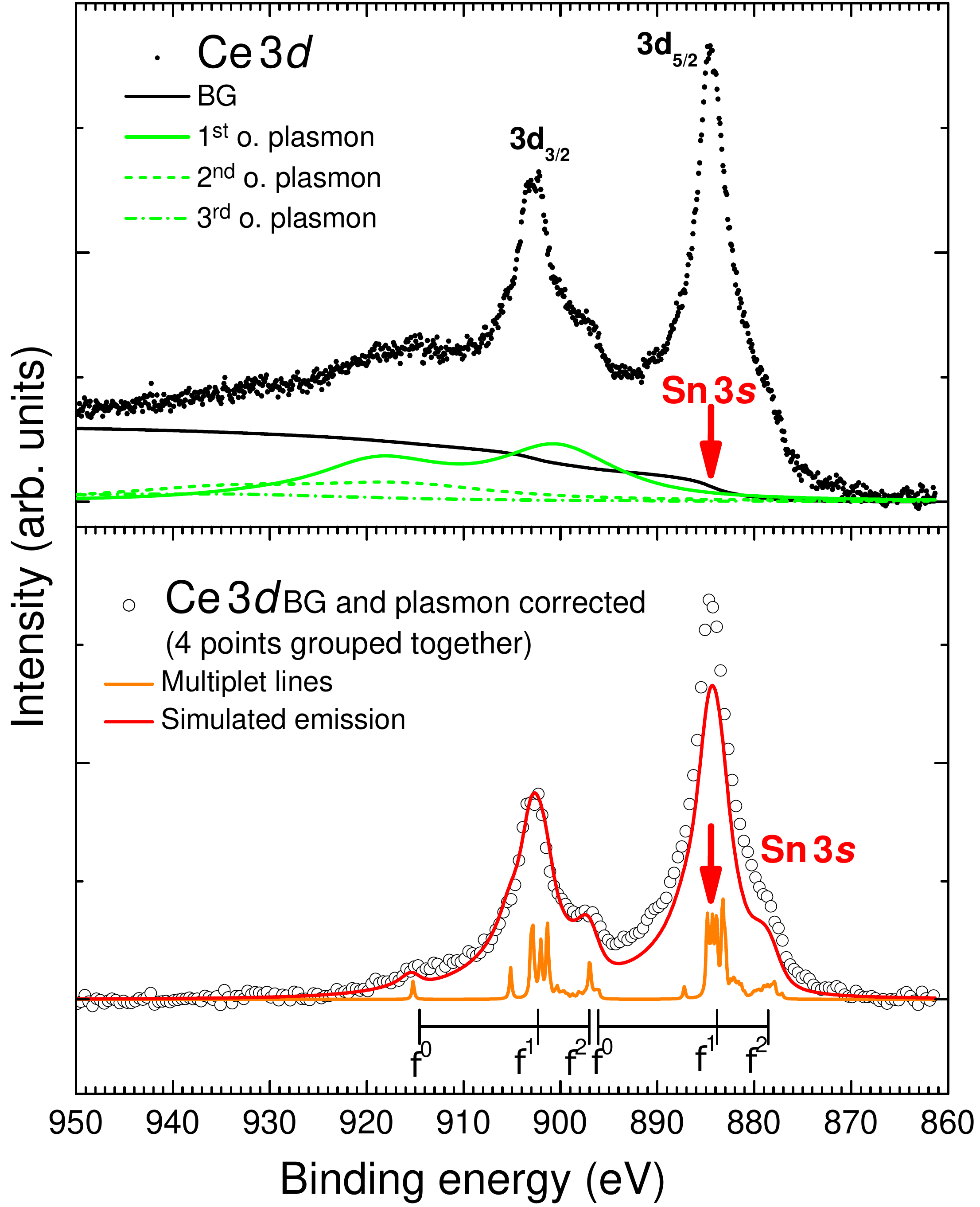}
\caption{\label{Ce3d}HAXPES data of Ce\,3$d$ core level emission. Top: data (black dots), Shirley-type background (black line), and plasmon intensities (green lines). Bottom: consistency check of the fit to Ce\,3$p$ data. in Fig.\,\ref{Ce3p}. Background and plasmon corrected data (black circles), multiplet structure from configuration interaction model with parameters as obtained from fit to Ce\,3$p$ in Fig.\,\ref{Ce3p} (orange line) and total calculated Ce\,3$d$ emission intensity (red line).  The discrepancy is due to the Sn\,3$s$. The bottom ruler indicates the position of the emission for the different 4$f$ occupancies.}
\end{figure}

\begin{figure}[h]
\centering
\includegraphics[height=26pc]{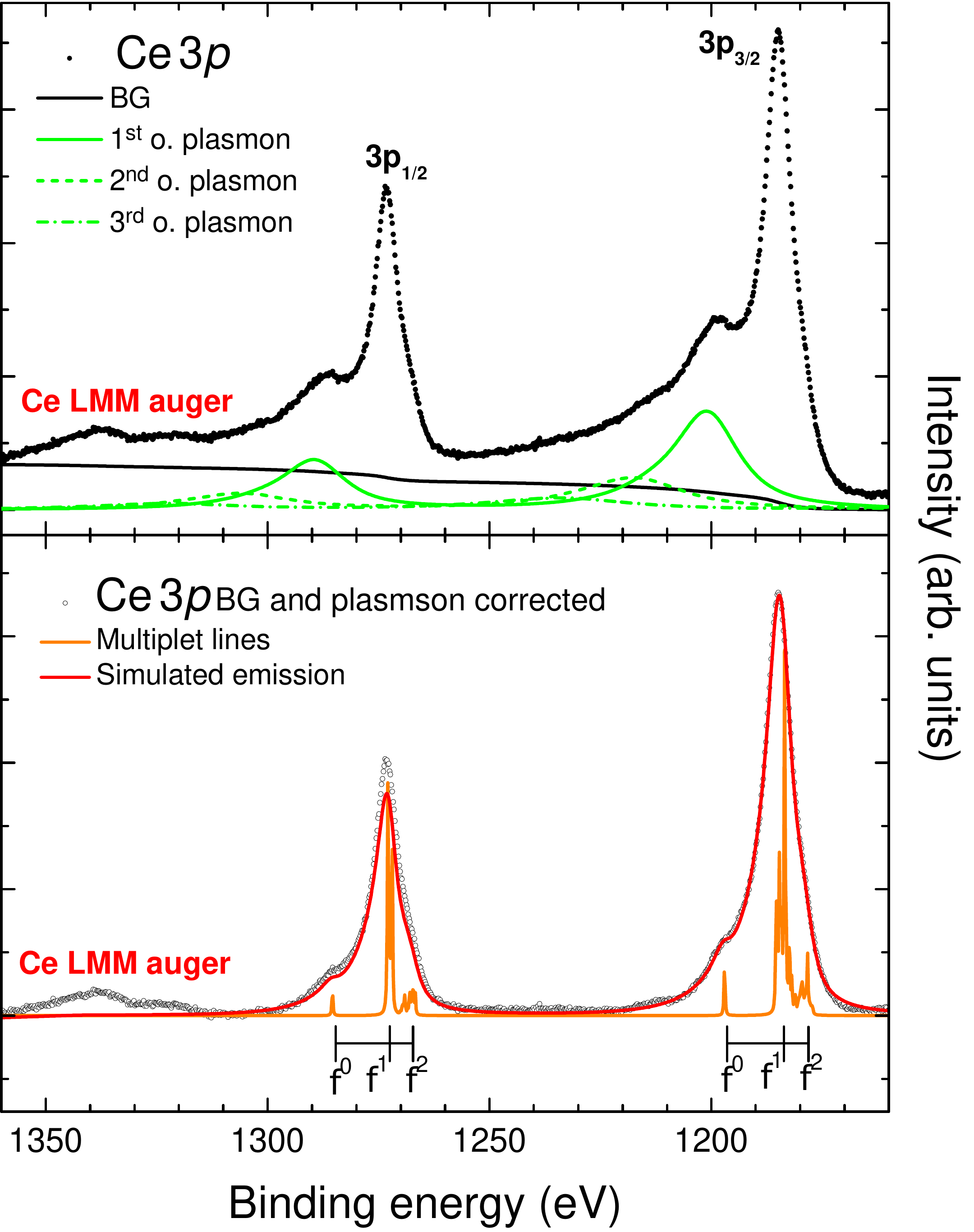}
\caption{\label{Ce3p}HAXPES data of Ce\,3$p$ core level emission. Top: data (black dots), Shirley-type background (black line), and plasmon intensities (green lines). Bottom: fit to background and plasmon corrected data (black circles), multiplet structure from optimized configuration interaction model (orange line) and total calculated Ce\,3$p$ emission intensity. The bottom ruler indicates the position of the emission for the different 4$f$ occupancies.}
\end{figure}

Single crystals were grown by the self-flux floating-zone melting method using optical heating in a four-mirror furnace\,\cite{Prokofiev2012}. The samples are single grain and single phase, as shown by Laue and x-ray diffraction, scanning electron microscopy, and energy dispersive x-ray spectroscopy.

Hard x-ray core level photoemission spectra are performed at the Taiwan beamline BL12XU at the SPring-8 synchrotron radiation facility in Japan.
The high incident energy of 6.47\,keV ensures bulk sensitivity.
The photo electrons were detected by a MB Scientific A-1HE analyzer with a pass energy of 200\,meV in vertical geometry for the Ce\,3$d$ test measurement and in horizontal geometry\,\cite{Weinen2016} for the Ce\,3$p$ and Sn\,3$p$, yielding an instrumental FWHM of about 1\,eV.
The Fermi energy was determined by a Au thin film measured with same conditions.
The samples were cooled down and cleaved in ultra high vacuum of $\approx$10$^{-8}$\,mbar and were then immediately transferred to the measurement position (pressure $<$3$\cdot$10$^{-10}$\,mbar and T$<$90\,K).
The quality of the sample and sample surface was verified by performing long energy scans that verify the absence of extra peaks (Fermi energy to 1500\,eV binding energy).

We have analyzed the HAXPES data with a combination of full multiplet and configuration interaction model. We use the XTLS~9.0 code\,\cite{TanakaJPSC63} for these calculations with input parameters from atomic calculations using Cowans atomic code\,\cite{Cowan}.
The code uses a configuration interaction model based on a single non-dispersive valence state\,\cite{Imer1987} combined with a full multiplet calculation.
The plasmon parameters are determined from emission lines that are not subject to the configuration interaction, here the Sn\,3$p$, so that, once the overall lineshape of the multiplet is adjusted, the configuration interaction model only depends on four parameters, Coulomb exchange interaction between the $f$ electrons ($U_{ff}$) and between the $f$ electrons and core hole ($U_{fc}$), the effective $f$-electron binding energy $\epsilon$$_f$ and the isotropic hybridization $V_\text{eff}$.
These parameters are uniquely defined by the intensity ratios and energy differences of the three valence states in the spectra. More details about the simulation can be found in Ref.\,\cite{Strigari2015} and \cite{Sundermann2016}.

\section{Results}

\begin{figure}[h]
\centering
\includegraphics[width=1.00\columnwidth]{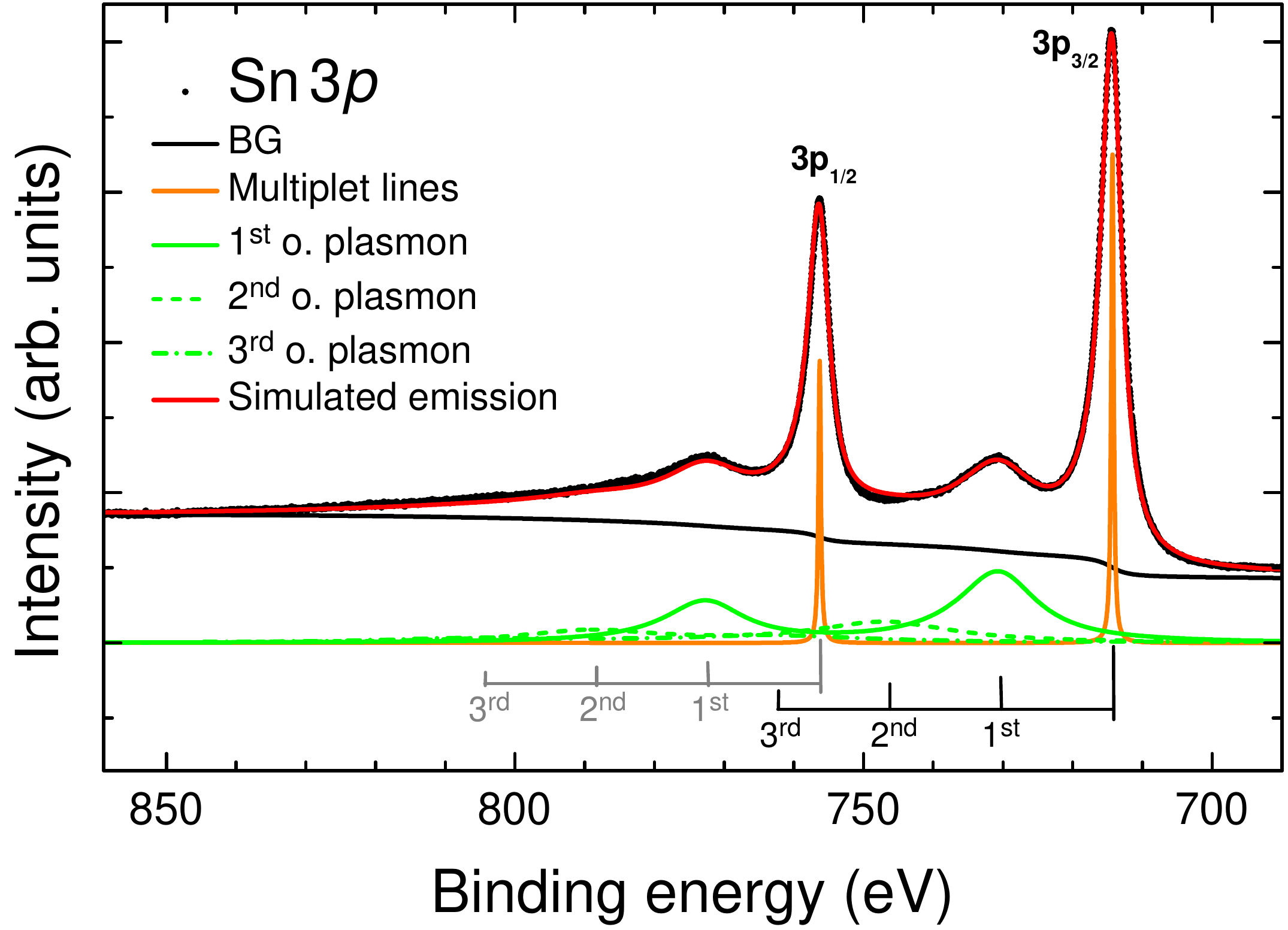}\hspace{2pc}%
\caption{\label{Sn3p}HAXPES data of the Sn\,3$p$ core level emission. Data (black dots), Shirley-type background (black line), main emission lines (orange line) and fitted plasmon intensities (green lines). The red line is the sum of the calculated contributions.}
\end{figure}

For the rare earth the 3$d$ core level spectrum is typically used for determining the valence state.
In Fig.\,\ref{Ce3d}\,(top) the Ce\,3$d$ spectrum of CeRu$_4$Sn$_6$ is shown.
There are two sets of emission lines due to the 3$d$ spin-orbit splitting (3$d_{3/2}$ and 3$d_{5/2}$). However, in CeRu$_4$Sn$_6$ the Sn\,3$s$ emission appears at 885\,eV binding energy (see red arrow) so that it severely distorts the spectral weights.
We therefore measured the Ce\,3$p$ emission line which is shown in Fig.\,\ref{Ce3p}\,(top).
The advantage of the Ce\,3$p$ is also that the two spin-orbit contributions do not overlap because the spin-orbit coupling of the Ce\,3$p$ electrons is much stronger and the resulting splitting of the Ce\,3$p_{1/2}$ and Ce\,3$p_{3/2}$ much larger than for Ce\,3$d$ (note the different energy scales of Fig.\,\ref{Ce3d} and \ref{Ce3p}). A drawback is the stronger intrinsic lifetime broadening due to the higher binding energy. The expected averaged energy positions of the $f^n$ configurations are marked by the ruler at the bottom of Fig.\,\ref{Ce3d} and \ref{Ce3p}, respectively.  

Before analyzing the Ce data further we look at the Sn\,3$p$ emission lines which are also spin orbit split (3$p_{1/2}$ and 3$p_{3/2}$) (see Fig.\,\ref{Sn3p}\,(top)). The multiplet structure consists of a single emission line and Sn is not affected by the configuration interaction so that the additional humps at about 16\,eV distance from the main emission lines are identified as strong plasmons. These plasmons have about the same energy distance from the main emission line as the expected $f^0$ contributions so that these plasmon intensities have to be taken into account when analyzing the Ce\,3$p$ spectra. After subtracting a Shirley-type background the plasmon parameters (energy positions, intensity ratios, and widths of 1st, 2nd, and 3rd order plasmons) are determined form the Sn\,3$p$ emission. In Fig.\,\ref{Sn3p}\,(top) the black line shows the Shirley-type background and the green lines the plasmon contributions (solid 1$^\text{st}$ order, dashed 2$^\text{nd}$ order, and dotted 3$^\text{rd}$ order). The orange line is the multiplet line broadened with a Gaussian FWHM=1\,eV and Lorentzian FWHM=3.1\,eV as well as a Mahan lineshape with energy cut-off $\xi$=8\,eV and asymmetry factor $\alpha$=0.1, and the red line is the total fit. The plasmon parameters are listed in Table\,\ref{tab_plasmon}.

We now use these plasmon parameters for describing the Ce\,3$p$ multiplet structure i.e.\ we attach the plasmon of this shape and energy to each emission line\,\cite{Strigari2015, Sundermann2016}. In Fig.\,\ref{Ce3p}\,(top) the Shirley-type background (black) and resulting plasmon intensities (green) are shown. In the bottom part the background and plasmon corrected data are shown. There is clearly a strong shoulder at the energy position where we expect the $f^0$ spectral weight. The combination of multiplet and configuration interaction calculation yields the multiplet structure (orange). For the fit the multiplet emission lines are broadened with a Gaussian FWHM=1\,eV, a Lorentzian FWHM=3.8\,eV and a Mahan lineshape with energy cut-off $\xi$=8\,eV and asymmetry factor $\alpha$=0.4. The red line is the total fit. The resulting weights $w_n$ of the $f^n$ configurations in the initial state and the resulting configuration interaction parameters are listed in Table\,\ref{tab_results}. The resulting $f^0$ weight ($w_0$) to the ground state amounts to 8\%. The strong Lorentzian broadening and overlap between $f^1$ and $f^2$ limits the precise separation between theses contribution so that we give the sum of $w_1$ and $w_2$. When fitting the main emphasis has been put on the Ce\,3$p_{3/2}$ part of the data since it is not affected by the Auger intensities.

As a consistency check the parameters from the Ce\,3$p$ emission data were used for Ce\,3$d$. The black and green lines in Fig.\,\ref{Ce3d}\,(top) refer to the Shirley-type background and the plasmon intensities, using the parameters from Table\,\ref{tab_plasmon}.
The bottom panel of Fig.\,\ref{Ce3d} shows the background an plasmon corrected data (4 energy intervals grouped together), the multiplet structure (orange) resulting from the multiplet calculation with the configuration interactions parameters from Table\,\ref{tab_results} and the calculated emission (red line).
The multiplet lines of the Ce\,3$d$ are broadened with a Gaussian FWHM=1\,eV, a Lorentzian FWHM=1.6\,eV and a Mahan lineshape with energy cut-off $\xi$=8\,eV and asymmetry factor $\alpha$=0.3.
The Ce\,3$d_{3/2}$ part is quite well described while the Ce\,3$d_{5/2}$ shows some discrepancy between calculation and data. The latter was expected because of the contribution of Sn\,3$s$.
Note, that the Ce\,3$d_{3/2}$ is still affected by the plasmon of the Sn\,3$s$.

\begin{table}
  \caption{Properties of the plasmons in CeRu$_4$Sn$_6$.}
  \label{tab_plasmon}
  \begin{tabular*}{\columnwidth}{l@{\extracolsep{\fill}}rrrr}
    \hline
		\hline
    \rule{0pt}{3.2mm}Order of plasmon & 1 & 2 & 3 & 4 \\
    \hline
    \rule{0pt}{3.2mm}Scaling factor & 0.5 & 0.25 & 0.13 & 0.06 \\
    Energy shift (eV) & 16.1 & 32.2 & 48.3 & 64.4 \\
    Lorentzian FWHM (eV) & 10.0 & 20.0 & 30.0 & 40.0 \\
    \hline
  \end{tabular*}
\end{table}

\begin{table}
  \caption{Optimized configuration interaction parameters ($U_{ff}$, $U_{fc}$, $V_\text{eff}$, $\epsilon_f$) and resulting $f^n$ weights $w_n$ in the ground state in \%.}
  \label{tab_results}
  \begin{tabular*}{\columnwidth}{c@{\extracolsep{\fill}}ccccc}
    \hline
		\hline
		\vspace{2mm}
    \rule{0pt}{3.8mm}$U_{ff}$(eV) & $U_{fc}$(eV) & $\epsilon_f$(eV) & $V_\text{eff}$(eV) & $w_0$(\%) & $w_1$+$w_2$(\%) \\

	9.1(9) & 10.6(9) & -2.5(3) & 0.26(2) & 8(2) & 92(2) \\
    \hline
  \end{tabular*}
\end{table}

The present experiment and analysis of the Ce\,3$p$ core level emission data finds 8\% of $f^0$ configuration in the initial state. The value is slightly larger than in our previous $L$-edge absorption study\,\cite{Sundermann2015} since in the present analysis final state effects have been taken into account. Comparing results of the same type of analysis, we can state that hybridization and occupation of $f^0$ of CeRu$_4$Sn$_6$ are only slightly larger than in CeRu$_2$Si$_2$, a magnetically non-ordering, metallic heavy fermion compound. However, ground state symmetries, that also have an impact on the efficiency of hybridization are not taken into account in such a comparison. 

We summarize that the Ce\,3$p$ core level is an alternative to the Ce\,3$d$ for determining valencies with photoemission and the present work provides a more quantitative value for the absolute $f^0$ weight in the ground state of CeRu$_4$Sn$_6$, thus confirming its moderate intermediate valency.
 
\section*{Acknowledgment}
K.C. M.S. and A.S. are grateful for support from the German funding agency DFG through Project  600575.

\end{document}